\documentclass[aps,prl,reprint,twocolumn,superscriptaddress,showpacs]{revtex4-2}

\usepackage{graphicx}
\usepackage{mathrsfs}
\usepackage{bm}
\usepackage{amsmath}
\usepackage{dcolumn}
\usepackage{epstopdf}
\usepackage{dsfont}
\usepackage{amssymb}
\usepackage{tabularx}
\usepackage{array}
\usepackage{float}
\usepackage{color}
\usepackage{epstopdf}
\usepackage{mathrsfs}
\usepackage[colorlinks, linkcolor=blue,anchorcolor=blue,citecolor=blue,urlcolor=blue]{hyperref}
\usepackage{multirow}

\begin{document}
\title{Upper bound of a band complex}
\author{Si Li}
\email{sili@nwu.edu.cn}
\affiliation{School of Physics, Northwest University, Xi'an 710127, China}
\affiliation{Shaanxi Key Laboratory for Theoretical Physics Frontiers, Xi'an 710127, China}

\author{Zeying Zhang}
\email{zzy@mail.buct.edu.cn}
\affiliation{College of Mathematics and Physics, Beijing University of Chemical Technology, Beijing 100029 , China}
\affiliation{Research Laboratory for Quantum Materials, Singapore University of Technology and Design, Singapore 487372, Singapore}

\author{Xukun Feng}
\affiliation{Research Laboratory for Quantum Materials, Singapore University of Technology and Design, Singapore 487372, Singapore}

\author{Weikang Wu}
\affiliation{Key Laboratory for Liquid-Solid Structural Evolution and Processing of Materials,
Ministry of Education, Shandong University, Jinan 250061, China}
\affiliation{Research Laboratory for Quantum Materials, Singapore University of Technology and Design, Singapore 487372, Singapore}

\author{Zhi-Ming Yu}
\affiliation{ Key Lab of Advanced Optoelectronic Quantum Architecture and Measurement (MOE),
	Beijing Key Lab of Nanophotonics $\&$ Ultrafine Optoelectronic Systems,
	and School of Physics, Beijing Institute of Technology, Beijing 100081, China}

\author{Y. X. Zhao}
\affiliation{National Laboratory of Solid State Microstructures and Department of Physics, Nanjing University, Nanjing 210093, China}
\affiliation{Collaborative Innovation Center of Advanced Microstructures, Nanjing University, Nanjing 210093, China}

\author{Yugui Yao}
\affiliation{ Key Lab of Advanced Optoelectronic Quantum Architecture and Measurement (MOE),
	Beijing Key Lab of Nanophotonics $\&$ Ultrafine Optoelectronic Systems,
	and School of Physics, Beijing Institute of Technology, Beijing 100081, China}

\author{Shengyuan A. Yang}
\affiliation{Research Laboratory for Quantum Materials, Singapore University of Technology and Design, Singapore 487372, Singapore}

\begin{abstract}
Band structure for a crystal generally consists of connected components in energy-momentum space, known as band complexes.
Here, we explore a fundamental aspect regarding the maximal number of bands that can be accommodated in a single band complex.
We show that in principle a band complex can have no finite upper bound for certain space groups. It means infinitely many bands can entangle together, forming a connected pattern stable against symmetry-preserving perturbations. This is demonstrated by our developed inductive construction procedure, through which a given band complex can always be grown into a larger one by gluing a basic building block to it. As a by-product, we demonstrate the existence of arbitrarily large accordion type band structures containing $N_C=4n$ bands, with $n\in\mathbb{N}$.

\end{abstract}
\maketitle

\begin{figure}[tb]
	\includegraphics[width=7.8cm]{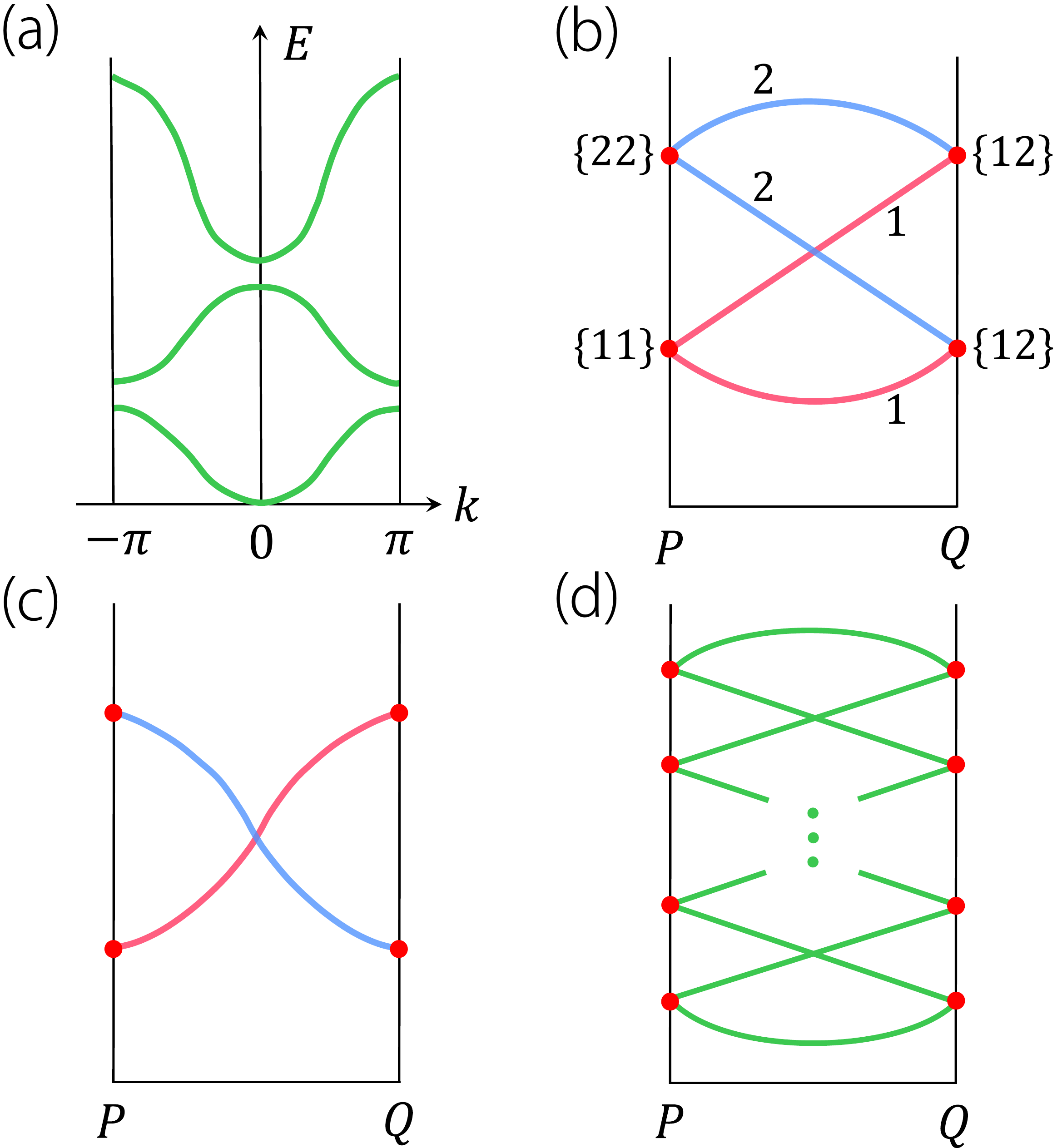}
	\caption{ (a) Band structure of 1D nearly-free particle model. (b) A band complex consisting of four bands. (c) These two bands do not form one complex, since they can be separated by shifting in energy. (d)  Schematic of an accordion type band complex.
		\label{fig1}}
\end{figure}

Band theory is of fundamental importance in condensed matter physics~\cite{ashcroft1976solid,cohen2016fundamentals}. Besides electrons and other quasiparticles in solid materials, band theory has also found great success in understanding various artificial crystal systems. In band theory, one considers a particle moving in a periodic lattice potential (which may include particle-particle interaction in a self-consistent way). Due to lattice periodicity, the particle wave vectors are restricted to the first Brillouin zone (BZ), which is the unit cell of reciprocal space, and the spectrum is composed of separated energy bands and band gaps.
For example, starting from a free particle model and imposing a weak lattice potential, one obtains what is commonly known as the nearly-free particle model~\cite{ashcroft1976solid}. The lattice potential $V$ introduces energy gaps $2|V_{\bm G}|$ at Bragg planes associated with the reciprocal lattice vector $\bm G$, with $V_{\bm G}$ the corresponding Fourier component of $V$. These gaps separate the original continuous spectrum into a series of bands. A typical band structure for the one-dimensional case is shown in Fig.~\ref{fig1}(a), where each band corresponds to a continuous curve in the energy-momentum space.

Crystalline symmetries may connect multiple bands together to form a band complex. For instance, along a high-symmetry path of BZ, denoted as $P$-$Q$, certain space groups can entangle four bands to form an hourglass pattern~\cite{young2015dirac,wang2016hourglass}, as in Fig.~\ref{fig1}(b). Clearly, this pattern requires the two middle bands belong to distinct representations of the little co-group on $P$-$Q$. A simplest case would be that: There are two one-dimensional (1d) representations on $P$-$Q$, labeled as 1 and 2; meanwhile, at high-symmetry points $P$ and $Q$, there exist 2d irreducible (co-)representations (IRRs), such that at $P$, the states correspond to the pairs $\{11\}$ and $\{22\}$, whereas at $Q$, they correspond to the pair $\{12\}$, as illustrated in Fig.~\ref{fig1}(b).
Viewed as a \emph{graph}, the pattern in Fig.~\ref{fig1}(b) consists of: ($a$) \emph{vertices} (the four dots), which correspond to protected degeneracies at high-symmetry points; and ($b$) \emph{edges} between vertices, which correspond to the band curves. We define such a \emph{connected} pattern as a {band complex}. Here, `connected' means that starting from a vertex, one can reach any other vertex via the edges in the complex. In this sense, Fig.~\ref{fig1}(b) is a band complex, but Fig.~\ref{fig1}(c) is not~\footnote{The pattern in Fig.~\ref{fig1}(c) can be easily untied by pushing one of the bands to higher energy.}.

The number of bands $N_C$ involved in a band complex $C$ is an important character. Recent studies established that for each space group (SG),  $N_C$ has a determined lower bound~\cite{parameswaran2013topological,watanabe2015filling,watanabe2016filling}. Again consider Fig.~\ref{fig1}(b). If a SG requires $\{12\}$ ($\{11\}$ and $\{22\}$) to be the only IRR(s) at $Q$ ($P$), then a band complex for this SG would be guaranteed to contain at least four bands, i.e., the lower bound of $N_C$ in this case $\geq 4$. The case achieving the lower bound of $N_C$ may be called a minimal band complex.

Now, a natural question is: \emph{Is there an upper bound on $N_C$?} In other words, what is the maximal number of bands that can be accommodated in a band complex? Interestingly, this fundamental question has never been explored before.

In this work, we show that at least from mathematical perspective, $N_C$ does \emph{not} have a finite upper bound for certain SGs. This is proved via
an explicit {construction} of a particular type of band complexes [illustrated in Fig.~\ref{fig1}(d)] that can reach $N_C\rightarrow \infty$ in certain SG. As a consequence, such a system can be metallic regardless of band filling.
It should be noted that the band complexes discussed here are \emph{stable} in the \emph{perturbative} sense, meaning that the pattern remains robust under symmetry-preserving perturbations to the system. However, a big change can still break it into parts, with each having a smaller $N_C$.

The band complex in Fig.~\ref{fig1}(d) can be viewed as a generalization of the hourglass in Fig.~\ref{fig1}(b). In some previous studies, this pattern was called an accordion band structure~\cite{zhang2018topological,zeng2020n,gatti2020radial,hirschmann2021symmetry,gonzalez2021chiralities}. However, in those works, the accordion was enforced to exist by some screw rotational symmetry, i.e., it represents a minimal band complex with its $N_C$ being a lower bound (can be up to 12 for sixfold screw axis~\cite{zhang2018topological}) required by symmetry.
In contrast, the complex here is not minimal (but it is stable as we emphasized) and it is not related to any screw symmetry. As a by-product of this work, we demonstrate the possibility of protected accordion band structures with \emph{arbitrarily large} $N_C=4n$, with $n\in \mathbb{N}$ being a natural number.

Our constructional proof proceeds in three steps. First, we propose a set of symmetry conditions to achieve the target
band complex in Fig.~\ref{fig1}(b). Under these conditions, we search and identify possible candidate SGs. Second,
for the candidate SG, we analyze its minimal band complexes. For each minimal complex, we show how it can be realized by a concrete lattice model. Third, we design an inductive glue procedure, such that starting from a complex $C$ with $N_C=4n$, we can always construct a larger complex $C'$ with $N_{C'}=4n+4$ by gluing a minimal complex. This inductive construction proves the absence of a finite upper bound.
\\
\begin{figure}[tbp]
	\includegraphics[width=8cm]{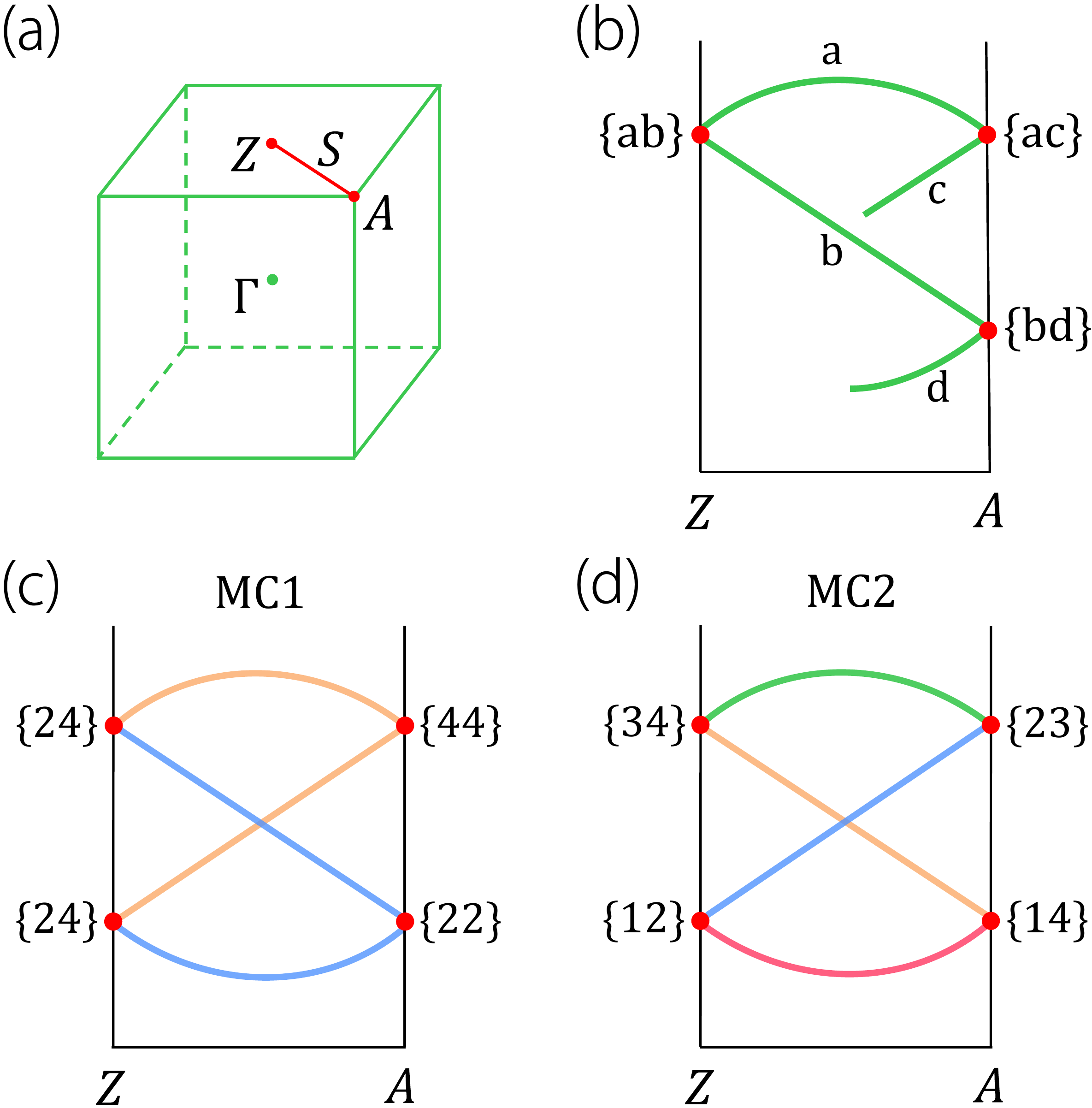}
	\caption{ (a) Brillouin zone for SG~138. We focus on $Z$-$A$ path. $S$ is a generic point on this path. (b) shows a band complex on $Z$-$A$ must have $N_C\geq 4$. (c) and (d) are two representative minimal band complexes on $Z$-$A$. The labeling of IRRs are given in Table I.
		\label{fig2}}
\end{figure}

\textit{\textcolor{blue}{Symmetry conditions and SG search.}}
As a first step, we lay out two symmetry conditions that can help to realize the type of band complex in Fig.~\ref{fig1}(d).

($i$) On the high-symmetry path, denoted as $P$-$Q$, there exist \emph{only} 1d IRRs. The number $p$ of different 1d IRRs must be larger than one, i.e., $p>1$ (otherwise, two band curves cannot cross without opening a gap on this path).

($ii$) At the two high-symmetry points $P$ and $Q$, there exist \emph{only} 2d IRRs, corresponding to the degree-$2$ vertices in Fig.~\ref{fig1}(d).

The two conditions together also require that: Any 2d IRR at $P$ (or $Q$) must split into two 1d IRRs on the path $P$-$Q$. This splitting follows the compatibility relations.

We will adopt these two conditions to search for suitable SGs. It must be noted that they are definitely {not} necessary conditions. For example, it is certainly possible to have IRRs with other dimensions at the high-symmetry points. Nevertheless, our aim here is \emph{not} to find out all SGs that allow the target band complexes, and as we shall see, adopting these conditions helps to simplify the analysis.

With the two conditions, we search through all 230 SGs with time reversal symmetry $T$, also known as the type-II SGs. We consider both spinless and spinful cases (i.e., both single- and double-valued representations). The procedure is straightforward: For each SG, we examine all high-symmetry paths of its BZ, and by using references on IRRs for space groups, we check whether there is any path satisfying our prescribed two conditions.

Surprisingly, it turns out that the two conditions are actually quite stringent. Among all 230 type-II SGs, we find that only SG No.~138 ($P4_2/ncm$) in the spinless case satisfies the conditions on its $Z$-$A$ path. In the following, we shall focus on this SG to demonstrate our construction.
\\
\begin{table}[t]
	\caption{\label{table1} IRRs at $A$ and $Z$ for SG~138. `HLG' denotes the Herring little group at $Z$ and $A$. $\Gamma \downarrow S$ gives the compatibility relations, with $S$ some generic point on path $Z$-$A$.}
	\begin{ruledtabular}
		\begin{tabular}{ccccc}		
			\quad Point & HLG &\qquad  IRR $\Gamma$  & \qquad Our Notation & \qquad $\Gamma \downarrow S$  \\
			\hline 	
			\quad \multirow{4}{*}{${Z}$} &\multirow{4}{*}{$G_{32}^2$} &\qquad $Z_1$  & \qquad $\left\{ 13 \right\}$ & \qquad $S_1 \oplus S_3$  \\
			& &\qquad $Z_2$  & \qquad $\left\{ 34 \right\}$ & \qquad $S_3 \oplus S_4$  \\
			& &\qquad  $Z_3$   &\qquad $\left\{ 24 \right\}$ & \qquad $S_2 \oplus S_4$  \\
			& &\qquad  $Z_4$   &\qquad $\left\{ 12 \right\}$ & \qquad $S_1 \oplus S_2$  \\
			\hline
			\quad \multirow{6}{*}{$\mathrm{A}$} &\multirow{6}{*}{$G_{32}^5$} &\qquad $A_1^{+} A_2^{+}$   & \qquad $\left\{ 22 \right\}$ & \qquad $S_2 \oplus S_2$  \\
		&	&\qquad $A_1^{-} A_2^{-}$    &\qquad $\left\{ 11 \right\}$ & \qquad $S_1 \oplus S_1$ \\
		&	&\qquad $A_3^{+} A_4^{+}$   &\qquad $\left\{ 33 \right\}$ & \qquad $S_3 \oplus S_3$  \\
		&	&\qquad $A_3^{-} A_4^{-}$   &\qquad $\left\{ 44 \right\}$ & \qquad $S_4 \oplus S_4$  \\
		&	&\qquad $A_5^{+}$   &\qquad $\left\{ 14 \right\}$ & \qquad $S_1 \oplus S_4$  \\
		&	&\qquad $A_5^{-}$   &\qquad $\left\{ 23 \right\}$ & \qquad $S_2 \oplus S_3$  \\
		\end{tabular}
	\end{ruledtabular}
\end{table}

\smallskip

\textit{\textcolor{blue}{Minimal band complexes.}}
Let's take a closer look at the candidate SG.
SG~138 is for a tetragonal crystal system. Its BZ is illustrated in Fig.~\ref{fig2}(a). As mentioned, we shall focus on $Z$-$A$ path which satisfies our prescribed conditions.

Here, the two high-symmetry points are located at $Z: (0, 0, \pi)$ and $A: (\pi, \pi, \pi)$. We shall denote points on $Z$-$A$ path by $S: (k,k,\pi)$ with $k\in (0,\pi)$. The little co-group is $4/mmm1'$ at points $Z$ and $A$, and it is $mmm'$ on path $Z$-$A$.

The single-valued IRRs at $Z$ and $A$ are listed in Table~\ref{table1}. Note that although the two points have the same little co-group,
their IRRs are different. This is because the group contains nonsymmorphic operations; for high-symmetry points on the BZ boundary, IRRs in the Bloch basis are obtained from representations for the central extension groups of the point group.
Here, owing to the different wave vectors at $Z$ and $A$, their central extension groups are different, resulting in different IRRs~\cite{bradley2009mathematical}.

\begin{figure}[tb]
	\includegraphics[width=8cm]{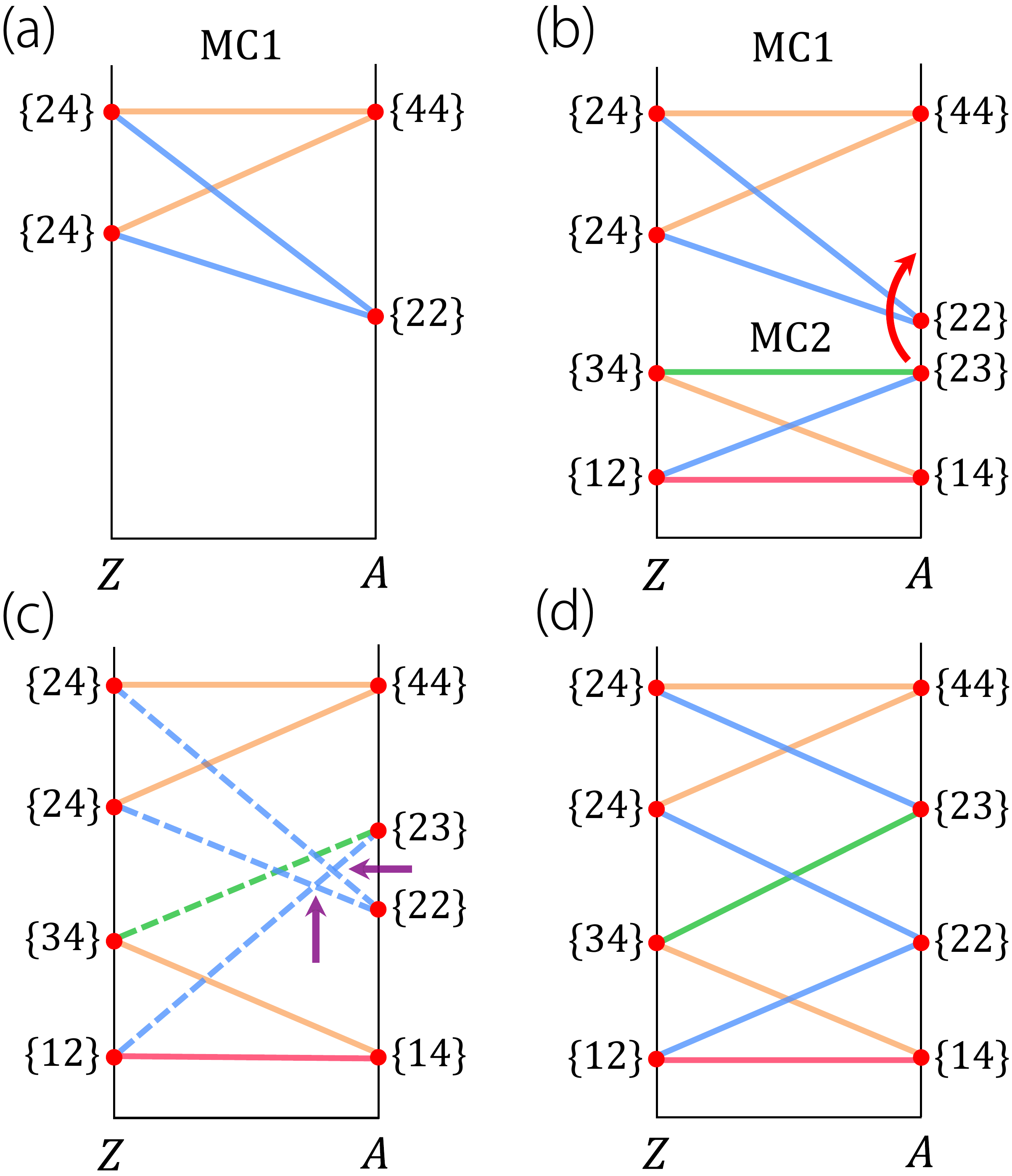}
	\caption{(a) Starting from a model for MC1, (b) one can bring in a model for MC2 and put it at energy below MC1. Fixing MC1 and tuning the MC2 model, one can always shift the $\{23\}$ vertex above $\{22\}$. (c) After switching orders, the two crossing points marked by purple arrows are not protected and should be gapped out, which leads to a single merged complex in (d).
		\label{fig3}}
\end{figure}

As required by our conditions, only 2d IRRs exist at $Z$ and $A$, and only 1d representations exist on $Z$-$A$.
In Table~\ref{table1}, we also give the compatibility relations, governing how each 2d IRR at $Z$ or $A$ split on $Z$-$A$.
One can see that there are totally four different 1d representations on $Z$-$A$, which we label as 1 to 4.
Without causing any ambiguity, we shall use the symbol $\{ab\}$ to label a 2d IRR at $Z$ or $A$, if it decomposes into representations $a$ and $b$ on $Z$-$A$ (with $a,b\in\{1,2,3,4\}$). The correspondence between the standard IRR notation and our labeling is also given in Table.~\ref{table1}.

Using Table~\ref{table1}, we can easily show that $N_C\geq 4$. To see this, let's start from an IRR $\{ab\}$ at $Z$. As a degree-2 vertex, it emits two edges  $a$ and $b$ to vertices at $A$. From Table~\ref{table1}, one observes that the IRR labels for $Z$ and $A$ share nothing in common. This means that: (1) The two edges $a$ and $b$ cannot connect to the same vertex at $A$; (2) the edge $a$ must connect to a vertex $\{ac\}$ at $A$ with $c\neq b$; and also edge $b$ must connect to a vertex $\{bd\}$ at $A$ with $d\neq a$. This situation is illustrated in Fig.~\ref{fig2}(b), from which one can see that a complex contains at least four bands.

Is the lower bound of $N_C$ equal to $4$? The answer is yes. Two minimal band complexes achieving $N_C=4$ are presented in Fig.~\ref{fig2}(c) and \ref{fig2}(d).
Note that these two are just representatives. We find that there are in total ten different minimal complexes for this path, all sharing the hourglass structure, which are given in Supplemental Material (SM)~\cite{SM}.

To show that these minimal complexes are physically possible, for each one, we show how it can be realized by a concrete lattice model. This step is facilitated by making use of the correspondence between real-space orbits and momentum-space band representations developed in Refs.~\cite{bradlyn2017topological,kruthoff2017topological,song2020twisted}, the information documented in the Bilbao database~\cite{aroyo2006bilbao1,aroyo2006bilbao2,aroyo2011crystallography}, and our developed MagneticTB package~\cite{zhang2022magnetictb,liu2021spacegroupirep,liu2023msgcorep}.
For example, the minimal complex in Fig.~\ref{fig2}(c) can be realized by putting one $p_z$-like orbital basis at $8h$ Wyckoff positions in a unit cell; whereas Fig.~\ref{fig2}(d) can be made by putting one $p_z$-like orbital basis at $4b$ Wyckoff positions. The details for these concrete models are given in SM~\cite{SM}. The conclusion is that: Each minimal complex here can indeed be realized by a simple lattice model.
\\

\smallskip

\textit{\textcolor{blue}{Inductive construction procedure.}} 
Now, we have things ready to construct the target band complex in Fig.~\ref{fig1}(d). The process follows the principle of mathematical induction.

First of all, we already have the target band complex with $N_C=4$, which is just the minimal complex in Fig.~\ref{fig2}(c) or Fig.~\ref{fig2}(d), and we have it realized by a lattice model. Actually, it turns out that our construction below only needs the two minimal complexes in Fig.~\ref{fig2}(c) and Fig.~\ref{fig2}(d). In the following, we shall refer to them as MC1 and MC2, respectively.

Starting from the complex with $N_C=4$, say MC1, we can construct a larger complex $C'$ with $N_{C'}=8$ by gluing a MC2 to it. Here is the procedure.

First, we can bring a MC2 at energy below the given MC1 [see Fig.~\ref{fig3}(a,b)]. Given that the two minimal complexes are resulting from two initially decoupled models,
this is easily achievable, e.g., one can simply tuning down the on-site energies in the lattice model for MC2.

Then, we glue MC2 to MC1 by switching the energy ordering at $A$ between the $\{22\}$ vertex of MC1 and the $\{23\}$ vertex of MC2 [see Fig.~\ref{fig3}(b,c)]. This can always be achieved. For example, one may fix the MC1 model and only tune the MC2 model, such that the switching at $A$ occurs and the energy ordering of vertices are as in Fig.~\ref{fig3}(c). Since the number of constraints is finite and we can in principle put as many
independent parameters in the MC2 model as we want, the switching can always be achieved in a finite parameter region.
A concrete model realization of this step is also given in SM~\cite{SM}.

After switching, clearly, only connection of the edges emitted from $\{22\}$ and $\{23\}$ vertices will get affected. As one can see from Fig.~\ref{fig3}(c), the two crossings between edges with the same IRR (the blue colored ones) on $Z$-$A$ must be gapped out. This results in the accordion pattern in Fig.~\ref{fig3}(d). In fact, one can easily convince him/herself that Fig.~\ref{fig3}(d) is the only possible connection pattern after switching. Now, we have successfully glued MC2 to MC1 and constructed a larger complex with $N_C=8$.

\begin{figure}[t]
	\includegraphics[width=8cm]{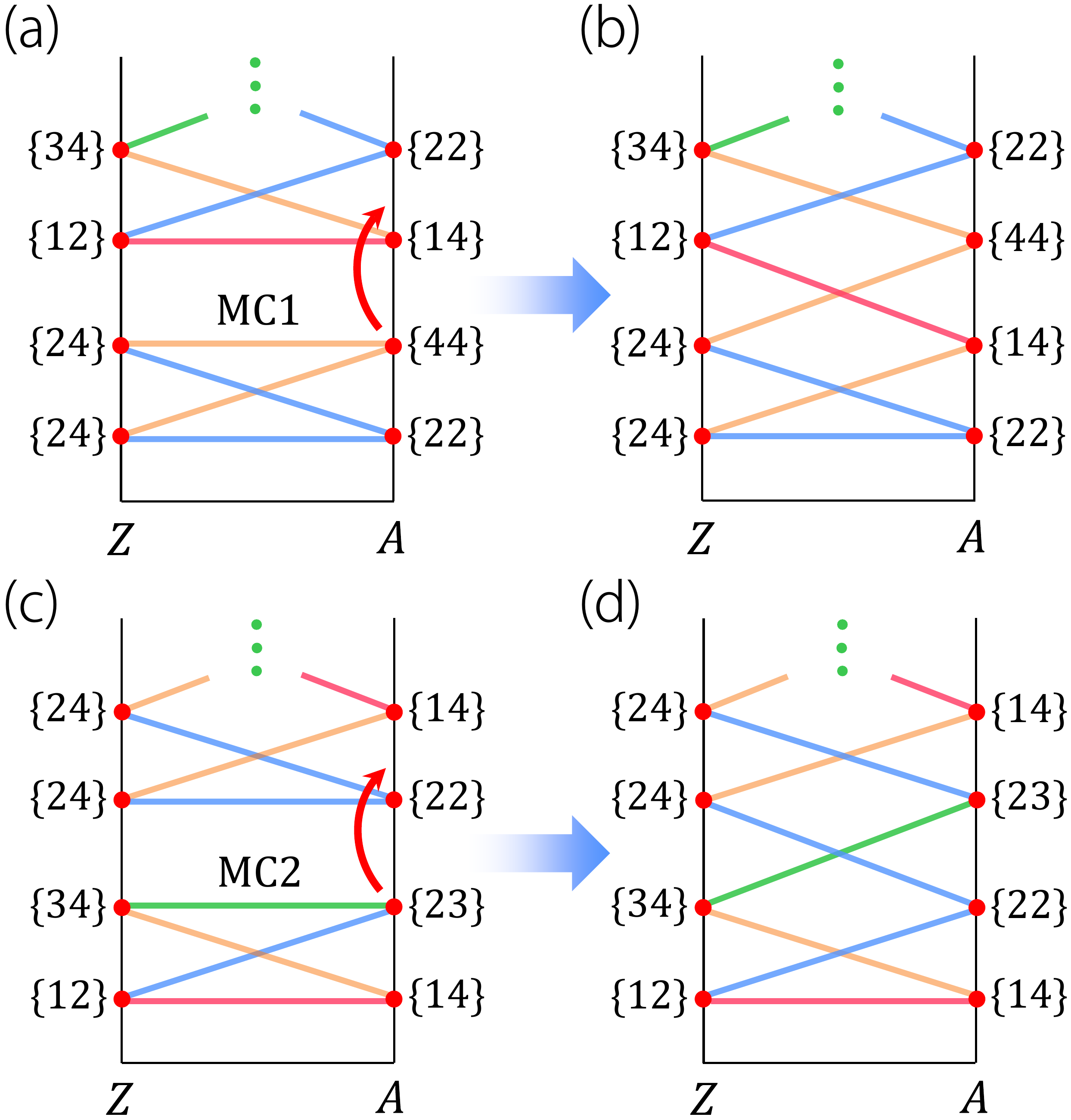}
	\caption{Inductive construction procedure. (a,b) When lowest vertices of a complex are the same as MC2, we can glue a MC1 to it. (c,d) After that, the lowest vertices are the same as MC1, we can then glue a MC2 to it. This process can continue indefinitely.
		\label{fig4}}
\end{figure}

Having understood the above glue procedure, the remaining is straightforward. Based on the $N_C=8$ complex in Fig.~\ref{fig3}(d), we can glue MC1 to it from below, as illustrated in Fig.~\ref{fig4}(a,b), by switching $\{14\}$ and $\{44\}$ at $A$. This gives a $N_C=12$ complex. Note that the lowest vertices in Fig.~\ref{fig4}(b) are the same as MC1, as it should be by our construction. Therefore, one can continue to glue a MC2 to it [see Fig.~\ref{fig4}(c,d)], and then MC1, and then MC2, ..., and this process can repeat indefinitely. In other words, after constructing a complex with $N_C=4n$, one can always grow it into a larger one with $N_{C}=4n+4$. Thus, we have proved our claim that we can have a case with no finite upper bound on $N_C$.

It must be noted that our constructed band complexes are indeed stable in the perturbative sense. For example, after forming the band complex in Fig.~\ref{fig3}(d), one may add all sorts of perturbations to the model, but as along as these perturbations are symmetry preserving and do not change the vertex ordering, the topology of the pattern must be maintained. This also means that
the constructed complexes occupy a finite region in the model parameter space, i.e., they indeed represent a phase of the system.
\\

\textit{\textcolor{blue}{Discussion.}}
Via an inductive construction procedure, we have proved an interesting result regarding the upper bound of a band complex. As a by-product, we showed that it is possible to have symmetry-protected accordion band structures with arbitrarily large $N_C=4n$. This is a new surprising result, in contrast to previous works where $N_C$ is limited to 12.

In our discussion, we have chosen the two minimal complexes in Fig.~\ref{fig2}(c,d) as basic building blocks. This choice is certainly not unique. As mentioned, there are other different minimal complexes. One can design different construction procedures by choosing different building blocks. An alternative construction approach is presented in SM~\cite{SM}.

Our construction is done for a particular SG, namely, the spinless type-II SG No.~138. Does the conclusion on upper bound apply to all SGs? The answer is no. It is clear that for the trivial SG, both the upper bound and the lower bound of $N_C$ would be 1. This indicates that the upper bound depends on SG, just like the lower bound.
Then the next question is to determine this value for each SG. It should be noted that: to show $N_C$ is unbounded,
it is sufficient to demonstrate it on a single high-symmetry path, as we did here. However, to study the upper bound of $N_C$ in general, one needs to consider all high-symmetry paths and points of BZ, which requires more effort.

Finally, we mention that our study demonstrated the existence of arbitrarily large accordion band complex in principle, but physically realizing it in a concrete physical system is a different question. For $N_C=8$, we find that it is realized in the phonon spectra of crystals AuCl and AuBr which belong to SG~138~\cite{SM}. Obviously, the larger the $N_C$ value is, the more difficult it is to find a real material realization. Nevertheless, the recent advance in artificial crystals has endowed us with
a great freedom to realize any kinds of lattice models~\cite{ozawa2019topological,ma2019topological,lu2014topological,yang2015topological,xue2020}. Particularly, electric circuit networks may be a good platform to
implement such models due to its versatility and high tunability~\cite{imhof2018topolectrical,lee2018topolectrical,yu20204d,wu2022non}.

~\

The authors thank D. L. Deng for helpful discussions. This work is supported by the NSF of China (Grant  No.~12204378, 12234003 and 12061131002), and the Singapore Ministry of Education AcRF Tier 2 (T2EP50220-0026).

\bibliography{ANL_ref}

\end{document}